\documentclass{emulateapj}
\usepackage{color}

\def\komkick{J0927+2943}
\def\j1050{J1050+3456}
\def\borobin{J1536+0441}

\def\etal{~et al.}
\def\simlt{\lower.5ex\hbox{$\; \buildrel < \over \sim \;$}}
\def\simgt{\lower.5ex\hbox{$\; \buildrel > \over \sim \;$}}

\def\gsim{\lower 2pt \hbox{$\, \buildrel {\scriptstyle >}\over
{\scriptstyle \sim}\,$}}
\def\lsim{\lower 2pt \hbox{$\, \buildrel {\scriptstyle <}\over
{\scriptstyle \sim}\,$}}

\def\deg{\ifmmode ^{\circ}
         \else $^{\circ}$\fi}
\def\pdeg{\ifmmode
           $\setbox0=\hbox{$^{\circ}$}\rlap{\hskip.11\wd0 .}$^{\circ}
     \else \setbox0=\hbox{$^{\circ}$}\rlap{\hskip.11\wd0 .}$^{\circ}$\fi}
     
\def\pc{\ifmmode \mathrm{pc} \else $\mathrm{pc}$ \fi}
\def\mpc{\ifmmode \mathrm{Mpc} \else $\mathrm{Mpc}$\fi}
\def\mpcthree{\ifmmode \mathrm{Mpc}^{-3} \else $\mathrm{Mpc}^{-3}$\fi}
\def\gpcthree{\ifmmode \mathrm{Gpc}^{-3} \else $\mathrm{Gpc}^{-3}$\fi}

\def\kelvin{\ifmmode \mathrm{K} \else {$\mathrm{K}$}\fi}
\def\kev{\ifmmode \mathrm{keV} \else $\mathrm{keV}$ \fi}

\def\lsun{\ifmmode {L_\odot} \else $L_\odot$\fi}
\def\msun{\ifmmode M_\odot \else $M_\odot$\fi}
\def\msunyr{\ifmmode M_\odot~\mathrm{yr}^{-1} \else $M_\odot~\mathrm{yr}^{-1}$\fi}

\def\cosi{\ifmmode {\cos\,i} \else $\cos\,i$\fi}

\def\zoiii{\ifmmode {\rm z_{\mathrm [O~III]}} \else $z_{\mathrm [O~III]}$\fi}

\def\heii{\ifmmode {\rm He{\sc ii}} \else He~{\sc ii}\fi}
\def\mgii{\ifmmode {\rm Mg{\sc ii}} \else Mg~{\sc ii}\fi}
\def\caii{\ifmmode {\rm Ca{\sc ii}} \else Ca~{\sc ii}\fi}
\def\ciii{\ifmmode {\rm C{\sc iii}]} \else C~{\sc iii}]\fi}
\def\civ{\ifmmode {\rm C{\sc iv}} \else C~{\sc iv}\fi}
\def\mgii{\ifmmode {\rm Mg{\sc ii}} \else Mg~{\sc ii}\fi}
\def\oi{\ifmmode {\rm [O{\sc i}]} \else [O~{\sc i}]\fi}
\def\oii{\ifmmode {\rm [O{\sc ii}]} \else [O~{\sc ii}]\fi}
\def\oiii{\ifmmode {\rm [O{\sc iii}]} \else [O~{\sc iii}]\fi}
 
\newcommand{\nii}{{\sc [N~ii]}}
\newcommand{\neiii}{{[Ne~{\sc iii}]}}
\newcommand{\nev}{{[Ne~{\sc v}]}}
\newcommand{\sii}{{\sc [S~ii]}}
\newcommand{\feii}{Fe~{\sc ii}}
\newcommand{\fevii}{[Fe~{\sc vii}]}

\def\teff{\ifmmode {T_{\rm eff}} \else $T_{\rm eff}$\fi}
\def\tmax{\ifmmode {T_{\rm max}} \else $T_{\rm max}$\fi}

\def\mbh{\ifmmode {M_{\rm BH}} \else $M_{\rm BH}$\fi}
\def\led{\ifmmode L_{\mathrm{Ed}} \else $L_{\mathrm{Ed}}$\fi}
\def\lbolflare{\ifmmode L_{\mathrm{bol,flare}} \else $L_{\mathrm{bol,flare}}$\fi}
\def\lagn{\ifmmode L_{\mathrm{agn}} \else $L_{\mathrm{agn}}$\fi}
\def\lbolagn{\ifmmode L_{\mathrm{bol,agn}} \else $L_{\mathrm{bol,agn}}$\fi}
\def\lbol{\ifmmode L_{\mathrm{bol}} \else $L_{\mathrm{bol}}$\fi}
\def\mdot{\ifmmode {\dot M} \else $\dot M$\fi}
\def\mdoto{\ifmmode {\dot{M}_0} \else  $\dot{M}_0$\fi}
\def\mdotf{\ifmmode {\dot{M}_\mathrm{flare}} \else  $\dot{M}_\mathrm{flare}$\fi}

\def\hnot{\ifmmode H_0 \else H$_0$ \fi}

\def\vkep{\ifmmode v_\mathrm{Kep} \else $v_\mathrm{Kep}$ \fi}
\def\vc{\ifmmode v_\mathrm{c} \else $v_\mathrm{c}$ \fi}

\def\vthree{\ifmmode v_{1000} \else $v_{1000}$ \fi}
\def\vrel{\ifmmode v_\mathrm{rel} \else $v_\mathrm{rel}$ \fi}
\def\vkick{\ifmmode v_\mathrm{kick} \else $v_\mathrm{kick}$ \fi}
\def\vkickz{\ifmmode v_{\mathrm{kick},z} \else $v_{\mathrm{kick},z} $ \fi}
\def\vkicky{\ifmmode v_{\mathrm{kick},y} \else $v_{\mathrm{kick},y} $ \fi}
\def\vchar{\ifmmode v_\mathrm{char} \else $v_\mathrm{char}$ \fi}
\def\eflare{\ifmmode E_\mathrm{flare} \else $E_\mathrm{flare}$ \fi}
\def\ekick{\ifmmode E_\mathrm{kick} \else $E_\mathrm{kick}$ \fi}
\def\ecoll{\ifmmode E_\mathrm{coll} \else $E_\mathrm{coll}$ \fi}
\def\ezero{\ifmmode E_\mathrm{0} \else $E_\mathrm{0}$ \fi}
\def\efac{\ifmmode \xi_\mathrm{E} \else $\xi_\mathrm{E}$ \fi}
\def\tqso{\ifmmode t_\mathrm{QSO} \else $t_\mathrm{QSO}$ \fi}
\def\tflare{\ifmmode t_\mathrm{flare} \else $t_\mathrm{flare}$ \fi}
\def\tzero{\ifmmode t_\mathrm{0} \else $t_\mathrm{0}$ \fi}
\def\tfac{\ifmmode \xi_\mathrm{t} \else $\xi_\mathrm{t}$ \fi}
\def\gfac{\ifmmode f_\mathrm{g} \else $f_\mathrm{g}$ \fi}
\def\lflare{\ifmmode L_\mathrm{flare} \else $L_\mathrm{flare}$ \fi}
\def\fflare{\ifmmode F_\mathrm{flare} \else $F_\mathrm{flare}$ \fi}
\def\nflare{\ifmmode N_\mathrm{flare} \else $N_\mathrm{flare}$ \fi}
\def\tshock{\ifmmode T_\mathrm{shock} \else $T_\mathrm{shock}$ \fi}
\def\rmin{\ifmmode R_\mathrm{1} \else $R_\mathrm{1}$ \fi}
\def\rmax{\ifmmode R_\mathrm{2} \else $R_\mathrm{2}$ \fi}
\def\rbound{\ifmmode R_\mathrm{b} \else $R_\mathrm{b}$ \fi}
\def\pbound{\ifmmode P_\mathrm{b} \else $P_\mathrm{b}$ \fi}
\def\mbound{\ifmmode M_\mathrm{b} \else $M_\mathrm{b}$ \fi}
\def\mbo{\ifmmode M_{\mathrm{b}0} \else $M_{\mathrm{b}0} $ \fi}
\def\ebo{\ifmmode E_{\mathrm{b}0} \else $E_{\mathrm{b}0} $ \fi}
\def\efinal{\ifmmode E_\mathrm{final} \else $E_\mathrm{final} $ \fi}
\def\tbound{\ifmmode t_\mathrm{b} \else $t_\mathrm{b}$ \fi}
\def\tagn{\ifmmode t_\mathrm{AGN} \else $t_\mathrm{AGN}$ \fi}
\def\rlim{\ifmmode R_\mathrm{lim} \else $R_\mathrm{lim}$ \fi}
\def\vlim{\ifmmode v_\mathrm{lim} \else $v_\mathrm{lim}$ \fi}
\def\vphi{\ifmmode v_\phi \else $v_\phi$ \fi}
\def\mlim{\ifmmode M_\mathrm{lim} \else $M_\mathrm{lim}$ \fi}
\def\tlim{\ifmmode t_\mathrm{lim} \else $t_\mathrm{lim}$ \fi}
\def\llim{\ifmmode L_\mathrm{lim} \else $L_\mathrm{lim}$ \fi}
\def\fqso{\ifmmode f_\mathrm{QSO} \else $f_\mathrm{QSO}$ \fi}

\def\hbeta{\ifmmode \rm{H}\beta \else H$\beta$\fi}
\def\hbetan{\ifmmode \rm{H}\beta_{\rm n} \else H$\beta_{\rm n}$\fi}
\def\hgamma{\ifmmode \rm{H}\gamma \else H$\gamma$\fi}
\def\hdelta{\ifmmode \rm{H}\delta \else H$\delta$\fi}
\def\hepsilon{\ifmmode \rm{H}\epsilon \else H$\epsilon$\fi}
\def\hzeta{\ifmmode \rm{H}\zeta \else H$\zeta$\fi}
\def\halpha{\ifmmode \rm{H}\alpha \else H$\alpha$\fi}
\def\lalpha{\ifmmode \rm{Ly}\alpha \else Ly$\alpha$}

\def\dvhb{\ifmmode \Delta v_{\hbeta} \else $\Delta v_{\hbeta}$\fi}
\def\dvmg{\ifmmode \Delta v_{\rm{Mg}} \else $\Delta v_{\rm{Mg}}$\fi}

\def\muobs{\ifmmode {\mu_{o}} \else  $\mu_{o}$ \fi}
\def\cosi{\ifmmode {\mathrm{cos}\,i} \else $\mathrm{cos}\,i$\fi}

\def\teff{\ifmmode {T_{eff}} \else $T_{eff}$ \fi}
\def\tmax{\ifmmode {T_{max}} \else $T_{max}$ \fi}

\def\tauh{\ifmmode {\tau_{\rm H}} \else $\tau_{\rm H}$ \fi}

\def\yr{\ifmmode {\rm yr} \else  yr \fi}
\def\kms{\ifmmode \rm km~s^{-1}\else $\rm km~s^{-1}$\fi}
\def\kmsyr{\ifmmode \rm km~s^{-1}~yr^{-1}\else $\rm km~s^{-1}~yr^{-1}$\fi}
\def\cm{\ifmmode {\rm cm} \else  cm \fi}
\def\cmmitwo{\ifmmode \rm cm^{-2} \else $\rm cm^{-2}$\fi}
\def\cmmithree{\ifmmode \rm cm^{-3} \else $\rm cm^{-3}$\fi}
\def\cmps{\ifmmode \rm cm~s^{-1}\else $\rm cm~s^{-1}$\fi}
\def\cmpsps{\ifmmode \rm cm~s^{-2}\else $\rm cm~s^{-2}$\fi}
\def\kmps{\ifmmode \rm km~s^{-1}\else $\rm km~s^{-1}$\fi}
\def\kmpspmpc{\ifmmode \rm km~s^{-1}~Mpc^{-1} \else
    $\rm km~s^{-1}~Mpc^{-1}$\fi}
  
\def\gcmthree{\ifmmode \rm g~cm^{-3} \else $\rm g~cm^{-3}$\fi}
\def\gcmtwo{\ifmmode \rm g~cm^{-2} \else $\rm g~cm^{-2}$\fi}
   
\def\erg{\ifmmode {\rm erg} \else $\rm erg$ \fi}
\def\ergps{\ifmmode {\rm erg~s^{-1}} \else $\rm erg~s^{-1}$ \fi}
\def\ergcms{\ifmmode \rm erg~cm^{-2}~s^{-1} \else $\rm erg~cm^{-2}~s^{-1}$ \fi}
\def\ergcmshz{\ifmmode \rm erg~s^{-1}~cm^{-2}~Hz^{-1} \else $\rm
erg~cm^{-2}~s^{-1}~Hz^{-1}$ \fi}
\def\ergcmsa{\ifmmode \rm erg~cm^{-2}~s^{-1}~\AA^{-1} \else $\rm
erg~cm^{-2}~s^{-1}~\AA^{-1}$ \fi}
\def\ergshz{\ifmmode \rm erg s^{-1} Hz^{-1} \else
   $\rm erg s^{-1} Hz^{-1}$ \fi}

\def\lam{\ifmmode {\lambda} \else {$\lambda$} \fi}
\def\llam{\ifmmode {L_\lambda} \else  $L_\lambda$ \fi}
\def\lamLlam{\ifmmode \lambda L_{\lambda}(5100) \else {$\lambda L_{\lambda}(5100)$} \fi}
\def\nuLnu{\ifmmode \nu L_{\nu}(5100) \else {$\nu L_{\nu}(5100)$} \fi}
\def\ilam{\ifmmode {I_\lambda} \else  $I_\lambda$ \fi}
\def\flam{\ifmmode {F_\lambda} \else  $F_\lambda$ \fi}
\def\inu{\ifmmode {I_\nu} \else  $I_\nu$ \fi}
\def\fnu{\ifmmode {F_\nu} \else  $F_\nu$ \fi}
\def\bnu{\ifmmode {B_\nu} \else  $B_\nu$ \fi}

\def\zthree{\ifmmode z_{\mathrm{[O~III]}} \else $z_{\mathrm{[O~III]}}$\fi}

\def\msigma{\ifmmode M_{\sigma} \else $M_{\sigma}$\fi}
\def\mbulge{\ifmmode M_{\mathrm{bulge}} \else $M_{\mathrm{bulge}}$\fi}
\def\mgal{\ifmmode M_{\mathrm{gal}} \else $M_{\mathrm{gal}}$\fi}
\def\lgal{\ifmmode L_{\mathrm{gal}} \else $L_{\mathrm{gal}}$\fi}
\def\lbulge{\ifmmode L_{\mathrm{bulge}} \else $L_{\mathrm{bulge}}$\fi}
\def\mgalstar{\ifmmode M^*_{\mathrm{gal}} \else $M^*_{\mathrm{gal}}$\fi}

\def\mbhsigstar{\ifmmode M_{\mathrm{BH}} - \sigma_* \else $M_{\mathrm{BH}} - \sigma_*$ \fi}
\def\deltalogmbh{\ifmmode \Delta~{\mathrm{log}}~M_{\mathrm{BH}} \else $\Delta$~log~$M_{\mathrm{BH}}$\fi}

\def\sigstar{\ifmmode \sigma_* \else $\sigma_*$\fi}
\def\sigthree{\ifmmode \sigma_{\mathrm{[O~III]}} \else $\sigma_{\mathrm{[O~III]}}$\fi}
\def\sigtwo{\ifmmode \sigma_{\mathrm{[O~II]}} \else $\sigma_{\mathrm{[O~II]}}$\fi}
\def\signl{\ifmmode \sigma_{\mathrm{NL}} \else $\sigma_{\mathrm{NL}}$\fi}
\def\wthree{\ifmmode {\rm FWHM({[O~III]})} \else $FWHM({[O~III]})$ \fi}
\def\wtwo{\ifmmode {\rm FWHM({[O~II]})} \else $FWHM({[O~II]})$ \fi}
\def\mthree{\ifmmode M_{\mathrm [O~III]} \else $M_{\mathrm [O~III]}$ \fi}
\def\mtwo{\ifmmode M_{\mathrm [O II]} \else $M_{\mathrm [O II]}$ \fi}

\def\lbreak{\ifmmode L_{\mathrm{break}} \else $L_{\mathrm{break}}$\fi}
\def\lcut{\ifmmode L_{\mathrm{cut}} \else $L_{\mathrm{cut}}$\fi}

\received{2009 July 20}
\accepted{2009 October 29}

\slugcomment{Astrophysical Journal 707 (2009) 936-941}

\shortauthors{Shields et al.}
\shorttitle{SDSS J1050+3456}

\begin{document}

\title{The Quasar SDSS~J105041.35+345631.3: Black Hole Recoil or Extreme Double-Peaked Emitter?}

\author{
G. A. Shields\altaffilmark{1}, 
D. J. Rosario  \altaffilmark{2}, 
K. L. Smith  \altaffilmark{1}, 
E.~W. Bonning \altaffilmark{3}, 
S.~Salviander \altaffilmark{1},
J. S. Kalirai  \altaffilmark{4},
R. Strickler  \altaffilmark{2}, 
E. Ramirez-Ruiz  \altaffilmark{2}, 
A. A. Dutton  \altaffilmark{2},
T. Treu \altaffilmark{5}, 
P. J. Marshall \altaffilmark{5}
}

\altaffiltext{1}{Department of Astronomy, University of Texas, Austin,
TX 78712, USA; shields@astro.as.utexas.edu, krista@mail.utexas.edu, triples@astro.as.utexas.edu} 

\altaffiltext{2}{Lick Observatory, University of California, Santa Cruz 95064, USA;
rosario@ucolick.org, rrs@ucolick.org, enrico@ucolick.org, dutton@ucolick,org}

\altaffiltext{3}{YCAA - Department of Physics, Yale University, New Haven, CT 06520, USA; erin.bonning@yale.edu}

\altaffiltext{4}{Space Telescope Science Institute, 3700 San Martin Drive, Baltimore MD, 21218, USA; jkalirai@stsci.edu}

\altaffiltext{5}{Department of Physics, University of California, Santa Barbara, 93106, USA; tt@physics.ucsb.edu, pjm@physics.ucsb.edu}

\begin{abstract}

The quasar SDSS J105041.35+345631.3 ($z = 0.272$) has broad emission lines 
blueshifted by $3500~\kms$ relative to the narrow lines and the host galaxy.  
Such an object may be a candidate for a recoiling supermassive black hole,  a binary black hole,
a superposition of two objects, or an unusual geometry for the broad emission-line region. The absence of narrow lines at the broad line redshift argues against superposition.  New Keck spectra of \j1050\  place tight constraints on the binary model.  The combination of large velocity shift and symmetrical \hbeta\ profile,  as well as aspects of the narrow line spectrum, make \j1050\  an interesting candidate for black hole recoil. Other aspects of the spectrum, however, suggest that the object is most likely an extreme case of a ``double-peaked emitter.''  We discuss possible observational tests to determine the true nature of this exceptional object.

\end{abstract}


\section{Introduction}
\label{sec:intro}

Mergers of binary black holes can result in the final black hole
receiving a `kick' due to anisotropic emission of gravitational
radiation at coalescence. Kick velocities are predicted to range up to
a maximum of 4000~\kms\ for certain configurations of black hole spin
and mass ratio \citep{campa07}. For the case of supermassive black
holes merging in a galaxy with sufficient gas to power an accretion
disk around the black hole, the merger remnant will carry a portion of
the disk with it and be visible as a quasar, either physically offset
from the nucleus or with broad emission lines Doppler-shifted from the
host galaxy velocity \citep{loeb07, madau04, blecha08}. Detection of a recoiling
black hole would be a powerful confirmation of the predictions of
numerical relativity and shed light on the cosmic evolution of
supermassive black holes.  Spectroscopic searches for kicked quasars 
 have had limited success.  \citet{bonning07}
reported upper limits to the incidence of kicks at various velocities in a sample of spectra
from the Sloan Digital Sky Survey \footnote{The SDSS Web site is
http://www.sdss.org.}.  One quasar, SDSS J092712.65+294344.0,
has been proposed by \citet{komossa08} as a
recoil candidate on the basis of its spectrum; however, other
physical models for this object have been proposed \citep{bogda09,
dotti09, heckman09, shields09}.   Another object, SDSS J153636.22+044127.0,  was proposed
as a sub-parsec binary supermassive black hole  on the basis of a large
blueshift of its broad emission lines \citep{boroson08}.  Subsequent observations  \citep{chornock09, lauer09} indicate that it is a member of the ``double-peaked emitter''  class of AGN, as discussed below.

The peculiar spectrum of  \j1050\ was noticed in the course of a search for binary QSO
candidates involving visual inspection of $\sim21,000$ spectra of QSOs at $0.1 < z < 0.7$
from SDSS Data Release 5 \citep{smith09}.  The object shows narrow
lines with typical AGN line ratios, at a redshift consistent with the stellar absorption lines.  The broad \hbeta\ line has a symmetrical profile  blueshifted by $3500~\kms$, near the upper end of the range predicted for gravitational radiation recoil.  In view of the importance of discovering examples of recoil in nature, all candidate objects merit close attention.  On the other hand, as emphasized by
\citet{bonning07} and \citet{chornock09}, most cases of displaced broad lines will be double-peaked emitters rather than kicks or binaries.  Here,
we present new spectroscopic observations of \j1050\ and discuss alternative models for its physical nature, including the possibility that it is an extreme case of a double peaked emitter.   We assume a concordance cosmology with H$_{0}=70$ \kms~Mpc$^{-1}$, 
$\Omega_{\Lambda}=0.7$. Spectral features are labeled using air wavelengths.

\section{Spectral Properties of \j1050}

\subsection{The SDSS Spectrum}

\begin{figure}[t]
\begin{center}
\includegraphics[angle=90, width=\columnwidth]{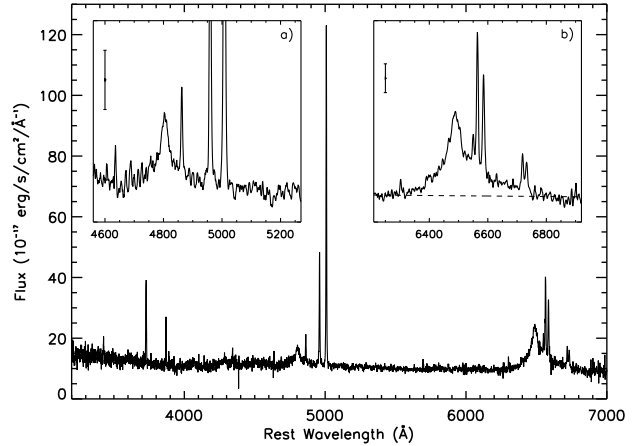}
\figcaption[]{
SDSS spectrum of \j1050\ plotted as observed flux versus rest wavelength.  
Insets show the regions of the \hbeta\ (left) and \halpha\ (right) lines.  
Note the blueshift of the peaks of the Balmer lines relative to the narrow lines, the symmetrical 
profile of \hbeta, and the extended red wing of \halpha.  In both insets, the vertical bar represents
$5\times10^{-17}~\ergcmsa$.
\label{fig:sdss} }
\end{center}
\end{figure}

The SDSS spectrum of \j1050\ is shown in Figure \ref{fig:sdss} and measurements are given in Table 1.   The spectrum shows strong, narrow emission lines including \oiii, \hbeta, \neiii, and \oii, and conspicuous stellar absorption lines.   The \oiii\ redshift is $\zthree = 0.2720$, and the other narrow emission lines as well as the stellar absorption lines are consistent within $\sim100~\kms$.
The intensity ratios among the narrow lines are typical of AGN.  Relative to \zthree,
the broad \hbeta\ line peaks at a blueshift of $3638\pm160~\kms$ and has a FWHM of $2170 \pm 250~\kms$. 
The peak of the broad \halpha\ line shows a similar blueshift ($3470\pm150~\kms$), but there is a more prominent red wing.  (Values in Table~\ref{t:tab1} for \mgii\ and the \nev\ flux and EW are from the Keck spectrum described below, with flux scaled to the SDSS spectrum using \neiii; other measurements are from the SDSS spectrum.)

The observed continuum flux at $5100(1+z)$~\AA\ is $\flam = 10.6\times10^{-17}~\ergcmsa$.   We have estimated the galaxy contribution to the continuum by modeling  the observed spectrum using a set  of empirical galaxy spectra from SDSS \citep{subbarao02}, combined with a simple power law with an adjustable spectral index for the AGN spectrum. The best fit is for an early-type galaxy.  We estimate the galaxy fraction of the observed continuum at rest wavelength 5100~\AA\ to be approximately 36\% with
an uncertainty of 0.1~dex.   For a typical early-type galaxy energy distribution, this implies an absolute magnitude $M_B = -20.0$ for the host galaxy.  For the local black hole--bulge relationship \citep[and references therein]{trem02} the expected \mbh\ is $10^{8.2\pm0.3}~\msun$, where the uncertainty reflects the scatter in the \mbh--bulge luminosity relationship.  Here we have
assumed that the observed stellar light is all bulge and made no allowance for fading.

The AGN component of the flux at $5100(1+z)$~\AA\ is $\flam = 7.4\times10^{-17}~\ergcmsa$, giving
$\lamLlam = 10^{44.13}~\ergps$.  This  implies a bolometric luminosity
$\lbol \approx 9\lamLlam = 10^{45.09}~\ergps$, following \citet{kaspi00}.
Objects with shifted broad lines may not conform to the usual relationship between
FWHM and black hole mass.  However, if we apply Equation~2  of \citet{shields03},
we find $\mbh(\hbeta) \approx 10^{7.5}~\msun$, using the measured \hbeta\ FWHM of
$2200~\kms$.  For this \mbh, the Eddington ratio of the AGN is $L/\led = 10^{-0.5}$, a typical value.

The FIRST survey \citep{becker95} detected no radio source at 20~cm at the position of \j1050
 down to a limit of 1~mJy.  This limit corresponds to a \citet{kellermann89} radio loudness parameter $\mathrm{log}\,R <  0.5$, assuming a radio spectral index $\fnu \propto \nu^{-0.5}$ between 6 and 20~cm.  Radio loud sources have $\mathrm{log}\, R > 2$.

\begin{deluxetable*}{lcccccccccccccc}
\tablewidth{\textwidth}
\tabletypesize{\scriptsize}
\tablecaption{Emission-Line Measurements\label{t:tab1}}
\tablehead{
\colhead{Ion }          &
\colhead{\sii}          &
\colhead{\sii}      &
\colhead{\nii} &
\colhead{\halpha(n)}         &
\colhead{\halpha(b)} &
\colhead{\oi}  &
\colhead{\oiii}    &
\colhead{\oiii}            &
\colhead{\hbeta(n)}            &
\colhead{\hbeta(b)}            &
\colhead{\neiii}      &
\colhead{\oii}  &
\colhead{\nev}  &
\colhead{\mgii(b)}
}
\startdata
$\lambda_{\mathrm rest}$ & 6730.8 & 6716.4 & 6583.4 & 6562.8 & 6562.8 & 6300.3 & 5006.8 & 4958.9 & 4861.4 & 4861.4 & 3869.1 & 3727.3 & 3425.9 & 2799.2 \\
$\lambda_{\mathrm obs}$ & 8561.8 & 8544.6 & 8374.3 & 8348.1 & 8254.5 & 8013.4 & 6368.6 & 6307.6 & 6183.3 & 6109.1 & 4920.7 & 4741.4 & 4357.2 & 3522.8 \\
$z$                               	& 0.2720 & 0.2722 & 0.2720 & 0.2720 & 0.2578 & 0.2719 & 0.2720 & 0.2720 & 0.2719 & 0.2570 & 0.2718 & 0.2720 & 0.2718 & 0.2585 \\
Flux	       & 60 & 72 & 180 & 230 &  2200 &  20.3 & 760 & 270 & 63 & 700 & 70 & 220 & 35 & 520 \\
Error & 6   & 7 & 18    & 12   &  300 &  5     &  38 &  13 & 6 & 70 & 4 & 11 & 7  & 80 \\
EW	      & 5.4 & 6.5 & 12.1 & 15.5 &  230 &  2.0 & 71 &  24 & 5.9 & 66 & 6.5 & 19 & 2.6 & 47 \\

\enddata

\tablecomments{Emission-line measurements for \j1050; ``b'' and ``'n'' denote peaks of the ``broad'' and ``narrow''  components.  Wavelength (air) and equivalent width are given in \AA; flux is in $10^{-17}~\ergcms$.  Flux and EW are in observed frame.}

\end{deluxetable*}

\subsection{Keck spectra}
\label{sec:keckspec}

\begin{figure}[t]
\begin{center}
\includegraphics[width=\columnwidth]{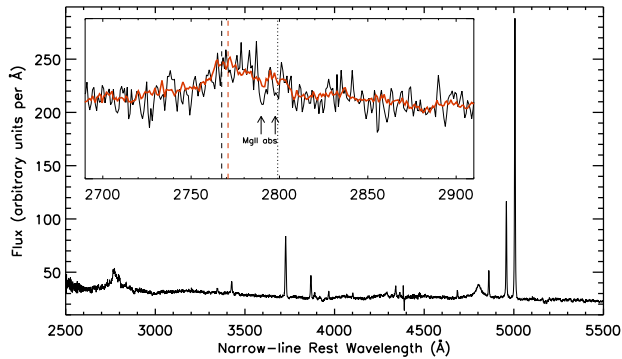}
\figcaption[]{
Keck LRIS spectrum of \j1050\ combining the 400g blue and 600g red spectra (2008 December).  Inset shows the region of the broad \mgii\ line, with the 400g spectrum in red and the higher resolution 600g spectrum in black.  Arrows mark the \mgii\ doublet with a blueshift of 500~\kms.  The black dashed line indicates the expected position of $\lambda2800$ for the redshift of the \hbeta\ broad line peak, 
while the red dashed line is the measured peak position of the \mgii\ broad line.
\label{fig:lris}
 }
\end{center}
\end{figure}

Blue and red spectra of \j1050\ were obtained on the morning of 2008 December 24 with the 
Low Resolution Imaging Spectrometer (LRIS)
double spectrograph on the Keck I  telescope \citep{oke95}. On the blue side, the 400/3400 grism 
(spectral resolution of 7.2~\AA\ FWHM) was used, which covered a wavelength region from the atmospheric cutoff to the dichroic at $\sim$5500~${\rm \AA}$.  On the red side, the 600/7500 grating (spectral resolution of 4.7~\AA\ FWHM), centered at 6600~${\rm \AA}$, covered a wavelength baseline of 2620~${\rm \AA}$.  The spectrum is shown in
Figure \ref{fig:lris} and measurements are given in Table 1.

One goal was to measure the wavelength and profile of
the \mgii~$\lambda\lambda2796, 2803$  line, noted by \citet{bonning07}
as a test of black hole recoil in cases of shifted Balmer lines.  The \mgii\ profile in Figure \ref{fig:lris} is 
irregular and the central wavelength somewhat ambiguous.  
Accordingly, we obtained a second LRIS spectrum, this time with a blue side 600/4000 grism  
(spectral resolution of 4.0~\AA\ FWHM), on the evening of 2009 April 9.  Though at much lower
signal-to-noise ratio (S/N), the improved spectral resolution of the newer blue spectrum
makes clear that the red wing of the broad \mgii\ feature is affected by a narrow \mgii\
absorption doublet at a blueshift of $500\pm50~\kms$ with respect to \zoiii.   
Because of the absorption doublet and limited S/N of the spectra, the line peak and centroid velocity of \mgii\ remain uncertain.   The \mgii\ peak in the 400g spectrum falls at $\sim -3151~\kms$, which is $490~\kms$ less than the \hbeta\ peak shift.   If this peak velocity is adopted for \mgii, then the profile has a substantial red wing, more pronounced than \halpha.   The \mgii\ centroid in the 400g spectrum has a blueshift $\sim2300\pm300~\kms$ relative to \zoiii, substantially less than the blueshift of the broad \hbeta\ line.  

A common pattern in normal AGN is for \nev$~\lambda3426$ to be 
wider than \oiii$~\lambda5007$.
\citet{bonning07} suggest that this should not occur for an ionizing source displaced 
from the galactic nucleus by  recoil.  The SDSS spectrum  lacks sufficient S/N for \nev.  
The 600/4000 LRIS spectrum shows a FWHM
for \nev\ of $5.4\pm0.3$~\AA, only slightly larger than the instrumental FWHM of 4~\AA.    Because of
this marginal resolution, and the possibility that the QSO image did not completely fill the spectrograph slit, we derived the effective spectral resolution by forcing the corrected FWHM of  \oii~$\lambda3727$ in the 600~$\mathrm{g~mm^{-1}}$ spectrum to agree with that of the SDSS  spectrum, which has better spectral resolution.  The result is a Keck instrumental FWHM of 3.6~\AA, yielding a corrected FWHM for \nev\ of $277\pm30~\kms$.  This  is close to the \oiii\  corrected FWHM of $273~\kms$ from the SDSS spectrum.

\section{The Nature of \j1050}
\label{sec:nature}

\subsection{Superposition in a Cluster?}
\label{sec:cluster}

Based on the SDSS field images around the QSO, as well as an examination of photometric redshifts 
of nearby bright galaxies, it is clear that \j1050\ is not in a rich cluster environment.
However, the nearest bright galaxy to the QSO does have a comparable photometric redshift, and has
colors and a derived absolute magnitude (M$_B \sim -21.7$) consistent with a massive early-type system. 
Since luminous red galaxies are generally found in relatively clustered environments \citep{zehavi05}, \j1050\
could possibly reside within a sparse cluster or galaxy group. Deeper imaging and spectroscopy of faint galaxies around the QSO may resolve this issue.

In a discussion of  \komkick, \citet{shields09} estimate a probability of order
$10^{-4}$ of superposition within 1~arcsec of two AGN with a velocity difference of order $2600~\kms$,
representing a chance occurrence in a rich cluster of galaxies.  Likewise, \citet{heckman09}
draw the analogy between \komkick\ and the radio galaxy
NGC 1275 (Per A) at the center of the Perseus cluster of galaxies, which has two narrow
emission-line systems offset in velocity by $\sim3000~\kms$. 
\j1050\ differs from \komkick\ and NGC~1275 in that it does not have narrow
emission lines at the peak velocity of the broad lines.  AGN rarely have
a total absence of narrow lines at the systemic velocity.  Weak narrow lines
tend to occur when permitted \feii\ is strong; however, \feii\ is weak in \j1050.  This
argues against a superposition as the explanation of the spectrum of \j1050.  A similar argument is 
given by \citet{lauer09} for \borobin.

\subsection{A Supermassive Black Hole Binary?}
\label{sec:binary}

Supermassive black hole binaries are expected as galaxy mergers lead to coalescing galactic nuclei
\citep{begelman80}. The orbital decay may stall at  a radius of 1~pc, comparable to the size of the BLR.  AGN with shifted or double-peaked broad lines have long been discussed as candidates for
binaries \citep{gaskell82}, but the expected changes of velocity shift with orbital phase have not been observed.   Recently, the binary interpretation has been offered for black hole recoil candidate \komkick\
\citep{bogda09,dotti09} and for the quasar SDSS \borobin\ \citep{boroson08, lauer09}.   
Like \j1050, these systems  have a broad \hbeta\ peak blueshifted by a several thousand \kms.   
\komkick\ has narrow lines at the velocity of the broad line peak,  supporting the suggestion that it might represent a superposition of two AGN \citep{heckman09, shields09}.   For \borobin, a strong case can be made for a double-peaked emitter \citep{chornock09, gaskell09}.  

Is the binary interpretation viable for \j1050?   The SDSS and Keck spectra together constrain the rate of change of the velocity shift of the \hbeta\ broad line peak relative to the narrow lines, a stable reference.  For the SDSS spectrum (dated 2005.17), the \hbeta\ shift is $\Delta v = -3638\pm 160~\kms$, relative to
the redshift of \oiii~$\lambda5007$, where the uncertainty is based on extreme cursor settings.  For the Keck spectrum (2008.98), we find $\Delta v = -3546\pm120~\kms$. This gives a rate of change of $dv/dt = +31\pm60~\kms$ per rest-frame year.  In the simple approximation that the AGN ($m_2$) is in circular orbit around $m_1$ in a plane containing the observer, the rate of change of line-of-sight velocity $u_2$ is
$
du_2/dt =  350 M_{1,8}^{-1} u_{2,3500}^4  s_\phi^4 c_\phi
$
in \kmsyr, where $M_{1,8} \equiv M_1/10^8~\msun$, $u_{2,3500} \equiv u_2/3500~\kms$, and
$s_\phi$ and  $c_\phi$ are the sine and cosine of the azimuth $\phi$.    For more detailed expressions (but with coefficients tailored to \komkick), see \citet{bogda09}.  If we interpret our measurement as an upper limit on $du_2/dt$ 
of 70~\kmsyr\ and take $m_1 = 10^{8.2}~\msun$ and $m_2 = 10^{7.5}~\msun$ (see Section~ \ref{sec:keckspec}), then $\phi$ is constrained to be within $\pm15~\deg$ of quadrature, a
fraction 17\% of the entire orbit.   For orbital inclination $i < 90\deg$  or for smaller masses, the constraint is tighter.
The combined mass of the black holes cannot be much larger, if the normal \mbh--bulge luminosity
relationship applies.

\subsection{A Recoiling Black Hole?}
\label{sec:recoil}

\j1050\ is in some ways an interesting candidate for recoil.  
(1)  The combination of large velocity shift
and relatively narrow symmetrical  profile, especially for \hbeta,  is just the signature proposed 
by \citet{bonning07} for recoil candidates.   
(2)  Our Keck spectrum shows a similar width for \nev\ and \oiii, another criterion suggested by Bonning et al. (see discussion in Section~\ref{sec:keckspec}).  
(3)  \citet{shields09} suggested that the narrow emission lines of refractory elements might be weak in recoiling QSOs.  A displaced ionizing source might preserve the order-of-magnitude depletions of Fe and other elements in the ISM, whereas these elements typically have normal gas-phase abundances in the NLR \citep[][and references therein]{nagao03}.    Nagao et al. give a mean value of 
I(\fevii $\lambda6087$)/I(\nev $\lambda3425) = 0.52$ for type I AGN (broad lines) and 0.34 for
type II AGN, whereas we find an upper limit  of 0.12 in \j1050.  Similarly, the composite Type II quasar spectrum of \citet{zakamska03} gives a value I(\mgii $\lambda2800$)/I($\hbeta) = 0.81$, whereas we find an upper limit of  0.16 for this ratio in the narrow line spectrum of \j1050.  These results suggest that depletion of Fe and Mg into grains in the NLR of \j1050\ may be more severe than in most AGN.

On the other hand, there are reasons for caution.
(1)  
\j1050\ shows characteristics of a double peaked emitter (see below).
(2)
\citet{shields09} estimated a probability of only $\sim10^{-4.9}$ for a line-of-sight kick velocity of $2650~\kms$ as observed for \komkick.  For \j1050, the shift of $3500~\kms$ is closer to the
theoretical maximum of $4000~\kms$, resulting in an even lower probability.    Note, however, that this argument is sensitive to the exact value of the maximum recoil velocity, currently based on extrapolation \citep{campa07}.  Also note that \citet{lauer09} argue that an ejection velocity of this magnitude is more likely to result from a three-body interaction than a gravitational radiation recoil event.
(3)
The \sii\ doublet is
marginally resolved with a typical line ratio of I$(\lambda6717)$/I$(\lambda6731) = 1.2 \pm 0.1$, implying an electron density $240\pm130~\cmmithree$ from the ``NEBULAR''  software
\footnote{http://stsdas.stsci.edu/nebular} \citep{shaw95} .  This resembles the situation for the r-system narrow lines of \komkick, and the photoionization argument of \citet{shields09} carries over to \j1050 (given the similar \oii/\oiii\ intensity ratio).
With allowance for the electron density and continuum luminosity of \j1050, the result is a distance of  $\sim 10^{0.3\pm0.2}$~kpc for the ionizing source from the narrow emission-line gas.   This indicates an age of $10^{5.5\pm0.2}~\yr$, contributing to the low probability of catching a recoil in this stage.
(4)
The narrow line spectrum of \j1050\ is in most regards quite normal and conforms to the pattern of strong \oiii\ with weak \feii.  The question of whether the ionizing continuum from a displaced QSO would produce normal AGN 
narrow-line ratios in the host galaxy ISM requires investigation \citep{komossa08}.

\subsection{A Double-Peaked Emitter?}
\label{sec:doublepeak}

Double peaked broad Balmer lines are sometimes seen, and a number of cases have been interpreted in terms of line emission from a Keplerian disk \citep[and references therein] {eracleous94}.  Occasionally, the broad Balmer lines may have a single peak shifted relative to the narrow lines.  Often in such cases, the profile is asymmetric, with the red wing more extended in the case of a blue-shifted peak, and vice versa, as discussed by \citet{bonning07}.  This pattern
is evident in the \halpha\  and \mgii\ profiles of \j1050 (Figures~1 and 2).
In addition, the \mgii\ centroid is $2200~\kms$ blueshifted, smaller than for \hbeta.  \citet{bonning07} noted a pattern for \mgii\ to have a smaller shift than \hbeta\ among their displaced-peak AGN.  
The 3500~\kms\ blueshift of the Balmer line peak of \j1050\ is large but not unprecedented for a double-peaked emitter.  In Table 3 of \citet{strateva03}, among 138 double-peaked emitters, the shift of the blue peak $W_{\mathrm b}$ ranges from $-6700$ to $+800~\kms$, with 21 objects blueshifted $3500~\kms$ or more.  However, the combination of the large shift and the dominance and symmetry of the blue peak of \j1050\ is exceptional.  In Strateva et al.,  FWHM for \halpha\ ranges from 
3700 to $20,700~\kms$;  and the centroid at half-maximum, FWHM$_c$, ranges from $-2200$ to $+1500~\kms$,  with only 5 objects having FWHM$_c$ more negative than $-1500~\kms$.  The combination of low FWHM of $2200~\kms$ and large FWHM$_c$ of $-3500~\kms$ for \j1050\  stands apart among known double-peaked emitters.  The reason for this is the weakness of the red wing and absence of a significant red peak, so that the width and centroid at half-maximum are determined by the blue peak alone.

Several studies have noted a tendency for the low ionization forbidden lines to be strong in
double-peaked AGN.  \j1050 departs from this pattern.
\citet{eracleous94}, working with radio loud objects,  found that the ratio
I([O~I]$\lambda6300$)/I([O~III]$\lambda5007$) has a median value of 0.14
for their ``disk-like emitters" but only 0.05 for other broad-line radio galaxies (BLRGs).
From Table 1, we measure this ratio to be $0.03\pm0.01$, less than half the
lowest value among the disklike emitters in Table 10 of \citet{eracleous94}.

The spectrum of \j1050\ bears a strong resemblance to that of  SDSS \borobin,  originally suggested as a parsec scale binary black hole by \citet{boroson08}.  \borobin\  has a broad \hbeta\ line peak blueshifted by $3500~\kms$,  similar to the case of \j1050.  Observations of \borobin\ by \citet{chornock09} and \citet{lauer09}  make a strong case for a double-peaked emitter.  Most significantly, the \halpha\ broad line, out of range of the SDSS spectrum considered by \citet{boroson08}, shows a red peak at a similar velocity offset to the stronger blue peak.   This example reinforces the case for caution in making exotic interpretations of \j1050\ and other objects with displaced broad line peaks.   The red--blue asymmetry in \j1050 is more extreme than for \borobin.  At a velocity of $v=+3500~\kms$ from the narrow line redshift
(symmetrical with the blue peak), the SDSS spectrum of \j1050\ shows a low red wing with little indication of a secondary peak or shoulder.  The height of the red wing above the continuum at this velocity is $0.18\pm0.04$ times the height of the blue peak.  The corresponding red/blue ratio in \borobin\ is $0.43\pm0.04$.  A further parallel between \j1050\ and \borobin\ is the presence of narrow \mgii\  absorption in both.  In \borobin, the \mgii\  absorption is at -240~\kms\ with respect to systemic velocity of the host galaxy, and in \j1050\ it is at  -500~\kms.  \citet{boroson08} argue that the geometry giving rise to the \mgii\ absorption may be difficult to explain in a model involving ejection of the black hole from the nucleus.  For \j1050, we have estimated that the recoiling black hole is about 2~kpc from the galactic center.  It seems possible that material ejected from the nucleus at an earlier stage of the AGN, left over from a recent galactic merger, or associated with a companion galaxy, might cause the \mgii\ absorption.

An alternative model of \borobin, proposed by \citet{tang09} , is a double-peaked emitter in a sub-parsec binary supermassive black hole system.  In their picture, there is a normal, fairly symmetrical double peaked profile from a Keplerian disk around the more massive hole, and the blue shifted peak comes from another BLR associated with the secondary black hole.  If such a model were applied to \j1050, the double-peaked disk emission would be considerably weaker relative to the blueshifted BLR emission from the secondary and would lack a well-defined peak on the red side of the line profile.

Could recoil (or three-body ejection) result in asymmetrical broad emission lines that mimic a
double peaked emitter?  The orbital period for the BLR is roughly $\sim10^2$~yr, so there should 
be ample time for a normal, symmetrical BLR to be re-established in the $\sim10^{5.5}$~yr  time since since the recoil event, as estimated above.  
This is the basis for rejection of asymmetrical broad line objects 
by \citet{bonning07}.  However, a better understanding of the dynamics of the retained accretion disk
and BLR in a kicked AGN is needed, including the longevity of asymmetries from the initial transient and the effect of a different outer boundary condition for a wandering AGN disk.

\section{Conclusion}
\label{sec:conclusion}

\j1050\ shows an unusual combination of a large velocity shift of the broad lines and
a fairly symmetrical profile, especially for \hbeta.  This object is in some respects an interesting
candidate for black hole recoil, in particular its highly blueshifted yet symmetrical broad \hbeta\ line and certain aspects of its narrow emission line spectrum.  Nevertheless, the presence of a weak but visible red wing on \halpha, and the lesser blueshift of the \mgii\ line, suggest that this object is most likely an extreme member of the class of double-peaked emitters.  Observations of the rest far-UV spectrum are needed to establish the velocity and profile of the high ionization broad lines, such as \civ, which tend to be narrower and less shifted than the Balmer lines in double-peaked emitters \citep{eracleous04, chornock09}.  Improved observations of the \mgii\ line profile are needed for a definitive measurement of the peak velocity and profile,
as well as the strength of narrow \mgii.  High resolution imaging can test for displacement of the AGN from the galactic nucleus and for tidal features indicative of a recent merger, as might be expected if the object is a recoil event or binary.   Multi-epoch spectroscopic monitoring can tighten constraints on the binary model and test for broad-line variability, as often occurs for double-peaked emitters.   If \j1050\ is confirmed as a double-peaked emitter, its extreme properties may lead to a better understanding of this class of AGN and help to develop criteria for distinguishing true cases of recoil or binary supermassive black holes.

\acknowledgments
G.S. acknowledges support from the Jane and Roland Blumberg
Centennial Professorship in Astronomy at the University of Texas at Austin.
Funding for the Sloan Digital Sky Survey (SDSS) has been provided by 
the Alfred P. Sloan Foundation, the Participating Institutions, NASA, NSF,  
the U.S. Department of Energy, the Japanese Monbukagakusho, and the 
Max Planck Society. The W.\ M.\ Keck
Observatory, which is operated as a scientific partnership among the
California Institute of Technology, the University of California, and the
National Aeronautics and Space Administration.  The Observatory was made
possible by the generous financial support of the W.\ M.\ Keck Foundation.

\end{document}